\documentclass[journal]{IEEEtran}
\usepackage{amsmath,amsfonts}
\usepackage{algorithmic}
\usepackage{array}
\usepackage[caption=false,font=normalsize,labelfont=sf,textfont=sf]{subfig}
\usepackage{textcomp}
\usepackage{stfloats}
\usepackage{url}
\usepackage{verbatim}
\usepackage{graphicx}
\usepackage{hyperref}
\usepackage{orcidlink}
\usepackage{booktabs}
\usepackage{multirow}
\usepackage{colortbl}
\usepackage{xcolor}
\usepackage{makecell}
\usepackage{tabu}
\renewcommand{\arraystretch}{1.5}
\newcolumntype{C}[1]{>{\centering\arraybackslash}m{#1}}

\usepackage[
    style=ieee,
    backend=biber,
    maxcitenames=1,
    mincitenames=1,
    maxbibnames=7,
]{biblatex}
\def\BibTeX{{\rm B\kern-.05em{\sc i\kern-.025em b}\kern-.08em
    T\kern-.1667em\lower.7ex\hbox{E}\kern-.125emX}}
\usepackage{balance}
\begin{document}
\title{FARM: Few-shot Adaptive Malware Family Classification under Concept Drift}

\author{
    Numan~Halit~Guldemir \orcidlink{0000-0003-1202-6841},
    Oluwafemi~Olukoya \orcidlink{0000-0003-2771-2553},
    and Jesús~Martínez-del-Rincón \orcidlink{0000-0002-9574-4138}%
    \thanks{The authors are with the Centre for Secure Information Technologies (CSIT), Queen’s University Belfast, United Kingdom. email: \href{mailto:nguldemir01@qub.ac.uk}{nguldemir01@qub.ac.uk}, \href{mailto:o.olukoya@qub.ac.uk}{o.olukoya@qub.ac.uk}, \href{mailto:j.martinez-del-rincon@qub.ac.uk}{j.martinez-del-rincon@qub.ac.uk}}
}

\maketitle
\begin{abstract}
Malware classification models often suffer performance degradation under concept drift due to evolving threat landscapes and the emergence of novel malware families. This paper presents FARM (Few-shot Adaptive Recognition of Malware), a unified framework for detecting and adapting to both covariate drift and label drift in Windows Portable Executable (PE) malware family classification. FARM uses a triplet autoencoder to project samples into a discriminative latent space, enabling unsupervised drift detection through DBSCAN clustering and dynamic thresholding. To enable rapid adaptation, the framework employs a few-shot strategy that can incorporate new classes from only a small number of labeled samples. FARM also supports full retraining when sufficient drifted samples accumulate, allowing longer-term model updating. Experiments on the BenchMFC dataset show that FARM improves classification performance under covariate drift by 5.6\%, and achieves an average F1 score of 0.85 on unseen malware families using few-shot adaptation, increasing to 0.94 after retraining. These results indicate that FARM provides an effective approach for drift-aware malware family classification in dynamic environments with limited supervision.
\end{abstract}


\section{Introduction}
\IEEEPARstart{T}{he} malware domain is inherently dynamic. Malware continuously evolves, either through the emergence of entirely new malware families with unknown characteristics or through the modification of existing malware behaviors by malware authors \cite{yang2021cade}. If malware detection and family attribution were based on a static view of the domain, these tasks might not appear particularly challenging. Indeed, leveraging machine learning and deep learning techniques, malware can often be detected with high accuracy when evaluated on static datasets. However, when faced with the real-world scenario of a constantly evolving malware landscape, the reliability and performance of malware detectors decrease significantly \cite{guldemir2024navigating, jordaney2017transcend, pendlebury2019tesseract}.

Empirical studies underscore the severity of this issue. For instance, when a malware classifier is trained on data collected at an initial time point and tested on future data, its detection performance deteriorates substantially, by up to 50\% compared to traditional randomized train-test splits that ignore temporal context \cite{pendlebury2019tesseract}. Similarly, another notable study indicates that within just six months of deployment, the performance of a malware detection model dropped from 99\% to 76\% \cite{chen2023continuous}. Such findings clearly illustrate the significant impact that evolving malware characteristics have on model effectiveness, emphasizing the necessity to address malware detection as a drift-aware adaptation problem, where models must detect distributional change and update their decision structure over time.

The observed performance degradation in malware detection models is commonly referred to as concept drift, also known as model aging \cite{xu2019droidevolver} or time decay \cite{pendlebury2019tesseract}. Concept drift arises when the statistical properties of data change over time, rendering previously effective models obsolete or less accurate \cite{lu2018learning}. Within the malware detection domain, concept drift manifests primarily in two distinct forms: label drift (inter-class drift \cite{he2024dream}) and covariate drift (intra-class drift \cite{he2024dream}).

Label drift describes the emergence of entirely new, previously unknown malware families, introducing novel patterns that detection models have never encountered. This form of drift challenges classifiers by presenting entirely new classes that require models to adapt or retrain to maintain their predictive capabilities. Covariate drift, on the other hand, refers to evolving behaviors and properties of existing malware families, where malware authors alter characteristics to evade detection while preserving the overall malware class identity \cite{jiang2024benchmfc, he2024dream}. Both types of drift necessitate ongoing adaptation or retraining of malware classifiers to sustain high detection accuracy and to effectively mitigate evolving cybersecurity threats.

The primary challenges in mitigating concept drift and enhancing malware family detection involve both model adaptation and data labeling. One key issue is updating the model to reflect evolving patterns while accurately determining whether a malware sample is novel or drifted. Relying solely on probability scores from machine learning or deep learning models is problematic, as these scores, especially from softmax-based classification layers, tend to remain high and overconfident even for previously unseen samples \cite{jordaney2017transcend, pearce2021understanding, hendrycks2016baseline}. Additionally, determining the optimal frequency for model retraining is challenging; too frequent updates may result in insufficient data for effective learning, while infrequent updates may fail to capture significant changes in malware behavior \cite{jordaney2017transcend}. Adding to these difficulties, labeling samples is expensive and time-consuming, particularly given the large datasets required by conventional deep learning approaches \cite{joyce2025claravy}. Moreover, labels provided by antivirus vendors or human experts are often unreliable, inconsistent, or prone to errors \cite{liang2021fare}.

Our approach addresses these challenges by first detecting potentially drifted samples in a learned metric space, using distances to family-level cluster representations rather than relying solely on classifier confidence. Samples that fall outside the accepted region of known families are treated as candidates for further analysis rather than being forcefully assigned to an existing class. To support rapid response under limited supervision, the framework then introduces a few-shot adaptation mechanism that can represent emerging malware behaviors using only a small number of labeled samples. When sufficient evidence accumulates, the model is retrained to consolidate these additions into the latent space. Accordingly, FARM is designed as a drift-aware adaptation pipeline with both immediate-response and longer-term consolidation stages.

While the majority of the prior work on concept drift and adaptation in malware detection has focused on the Android platform \cite{jordaney2017transcend, barbero2022transcending, yang2021cade, chen2023continuous, he2024dream}, our study focuses on Windows Portable Executable (PE) malware. According to industry-reported statistics, the daily detection volume of Windows PE malware exceeds 300,000 samples \cite{AVTest}. These incoming samples may correspond either to variants of existing malware families, which exhibit new behavioral patterns or characteristics, or to previously unseen malware families. Both cases can degrade the performance of static malware classifiers and reduce their reliability in real-world deployment.

Motivated by this challenge, we propose a unified framework for detecting and adapting to concept drift in Windows PE malware classification. Rather than treating drift detection and model adaptation as separate problems, our framework integrates both components to support rapid response to emerging samples and longer-term model updating. In summary, our work makes the following contributions:

\begin{itemize}
    \item We propose Few-shot Adaptive Recognition of Malware (FARM), a framework for Windows PE malware family classification under concept drift.
    \item We combine latent-space drift detection with few-shot adaptation to support the recognition of previously unseen malware families and evolving family variants.
    \item We evaluate the proposed framework on a timestamped malware benchmark under covariate-drift and label-drift scenarios, demonstrating its effectiveness in temporally evolving classification settings.
\end{itemize}

To structure our research, this paper is organized as follows. \autoref{sec:related-works} reviews related literature on concept drift in malware detection, metric learning, and few-shot adaptation. \autoref{sec:methodology} details the proposed FARM methodology, including the training, drift detection, and adaptation phases. \autoref{sec:eval} outlines the experimental setup. 
\autoref{sec:drift-detection} evaluates FARM’s ability to detect concept drift, including both covariate and label drift scenarios. \autoref{sec:drift-adaptation} focuses on adaptation performance, covering both few-shot integration and full retraining. \autoref{sec:discussion} discusses the broader implications, limitations, and potential improvements to our approach. Finally, \autoref{sec:conclusion} concludes the paper and outlines directions for future work.

    
\section{Related Work} \label{sec:related-works}

\subsection{Concept Drift in Malware Detection}
Concept drift refers to changes in data distribution over time, which can cause a model’s performance to degrade if not addressed. Various approaches have been proposed to address this problem by detecting drifting samples, understanding the underlying causes, and adapting the model accordingly.
\textcite{singh2012tracking} investigate concept drift in malware detection by proposing methods to measure how feature distributions change over time within malware families. Their study shows that while malware is generally expected to evolve, certain families exhibit limited drift, highlighting that not all detection models may require drift adaptation. \textcite{jordaney2017transcend} and \textcite{barbero2022transcending} use a conformal evaluator to assess incoming inputs and identify drifted samples, which are then excluded from immediate model training and earmarked for additional labeling. \textcite{yang2021cade} propose contrastive learning to map data samples into a low-dimensional space, using distance-based metric learning to select drifting samples and provide explanations for the observed drift. \textcite{chen2023continuous} builds on this idea by designing a custom loss function that encourages separation between stable and drifting representations. \textcite{he2024dream} proposed a concept-aware system for Android malware classification that improves drift detection and adaptation by embedding behavioral explanations into a contrastive autoencoder.


\subsection{Metric Learning}
Recent advances in metric learning have proven effective in developing high-quality data representations for tasks that involve differentiating objects and identifying semantic similarities across multiple domains, including computer vision \cite{hermans2017defense, dong2018triplet, mccartney2022zero}, audio processing \cite{oord2018representation, ghaleb2019metric, chung2020defence}, and bioinformatics \cite{xu2016multi, luo2019novel}. This area has also attracted considerable interest from security researchers because of its potential to improve malware detection. For instance, \textcite{wu2022contrastive} introduce IFDroid, a contrastive learning-based system for Android malware family classification. It converts function call graphs into images by applying centrality metrics from four different aspects, enabling the model to learn features via supervised contrastive learning. The system also employs a method to provide visual explanations for the predictions. Similarly, \textcite{jurevcek2021application} applies Particle Swarm Optimization (PSO) to adjust feature weights within a weighted Euclidean distance metric, thereby improving the performance of k-Nearest Neighbors (k-NN) classification. Additionally, \textcite{andresini2021autoencoder} propose a network intrusion detection method that combines autoencoders with triplet networks to enhance predictive accuracy by addressing data imbalance and improving the differentiation between normal and malicious network traffic. \textcite{rudd2024efficient} propose a metric learning framework for analyzing Windows PE malware by training embedding models on static features enriched with capability labels derived from disassembly. They use contrastive loss and a novel Spearman rank-based loss to capture both coarse and fine-grained similarities between samples, and show that a combination of these objectives improves performance.

\subsection{Few-shot Adaptation}

Meta-learning, or "learning-to-learn", trains models to quickly adapt to new tasks by leveraging experience from related ones \cite{hospedales2021meta}. Few-shot learning arises from this concept, enabling models to recognize new classes from just a few examples. A common few-shot approach is metric-based learning, which embeds data so that samples from the same class cluster together. \textcite{koch2015siamese} introduced Siamese networks for one-shot image recognition via a similarity function. Matching Networks \cite{vinyals2016matching} used attention over support examples (labeled examples per class) to classify queries (unlabeled test samples). Prototypical Networks \cite{snell2017prototypical} simplified this by computing class prototypes as averages of embedded features and classifying queries based on proximity to these prototypes. Few-shot learning is especially useful in cybersecurity, where labeled data for new threats is scarce. \textcite{rong2021umvd} propose a few-shot learning framework that uses Prototypical Networks \cite{snell2017prototypical} on grayscale images derived from network traffic data to detect unseen malware variants. \textcite{wang2021novel} propose a few-shot classification framework that leverages meta-learning and multiple prototypes to capture diverse behavioral patterns within malware families. By representing each family with a variable number of prototypes derived from dynamic analysis, their method improves robustness in recognizing previously unseen malware. \textcite{conti2022few} propose a few-shot malware classification framework based on Siamese networks and a novel GEM image representation (a 3-channel fusion of Markov images, entropy graphs, and gray-level co-occurrence matrices), achieving high accuracy using few samples per unseen malware family. \textcite{liu2022fewm} propose a few-shot malware detection approach that represents malware behavior as heterogeneous graphs. Using contrastive learning on these structured graphs, the model learns discriminative embeddings that enable accurate classification of novel malware variants with minimal labeled data.

While prior works have explored concept drift in malware detection, much of this research has focused on the Android platform \cite{jordaney2017transcend, barbero2022transcending, yang2021cade, chen2023continuous, he2024dream}. Existing approaches have primarily emphasized drift detection or continual updating, often without explicitly coupling drift identification with rapid adaptation and fine-grained malware family modeling \cite{yang2021cade, chen2023continuous}. Few-shot learning has also been studied in cybersecurity, but has been less explored in the context of temporally evolving Windows PE malware family classification. 

In this context, FARM is designed as a unified framework for Windows PE malware family classification under concept drift. It combines metric-learning-based drift detection with few-shot adaptation to support rapid incorporation of new or evolving malware families using only a small number of labeled samples. In addition, FARM integrates clustering and dynamic thresholding for drift identification together with a longer-term retraining strategy, enabling the framework to address both covariate drift and label drift within a single pipeline. We evaluate this design on the BenchMFC \cite{jiang2024benchmfc} dataset to study its effectiveness in temporally evolving malware classification settings.

\section{Methodology} \label{sec:methodology}

\begin{figure*}[ht]
    \centering
    \includegraphics[width=\linewidth]{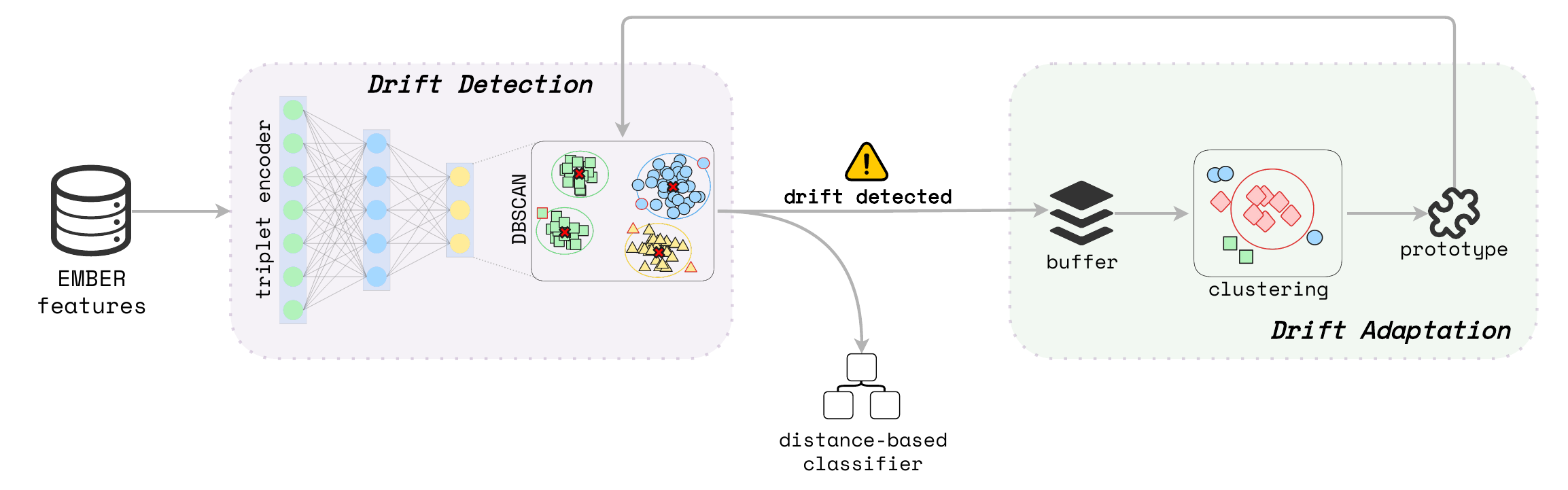}
    \caption{FARM inference and adaptation pipeline. The system detects concept drift in real-time, buffers uncertain samples, and dynamically adapts using few-shot prototype generation to maintain classification performance.}
    \label{fig:architecture}
\end{figure*}

Our approach involves multiple components, primarily the training of a triplet autoencoder network, which is utilized for both classification and concept drift detection, alongside a few-shot strategy to adapt the model to new data distributions. An overview of the FARM pipeline is illustrated in Figure~\ref{fig:architecture}. The system takes EMBER features \cite{anderson2018ember} as input, which are then embedded into a latent space using a triplet-trained encoder. In the \textit{drift detection} phase, DBSCAN is used to identify dense clusters and detect outliers or shifted samples. If a sample lies outside known cluster thresholds, it is flagged as drifted and passed to the \textit{drift adaptation} module. There, drifted samples are stored in a buffer and periodically clustered to discover coherent groups. Once a stable cluster is detected, a prototype is generated to represent a new or evolved malware family, allowing the classifier to integrate the new behavior.

\subsection{Training phase}
The triplet autoencoder model consists of two parts: an autoencoder and an embedding mechanism trained jointly using a combination of triplet loss and mean squared error (MSE) loss. The autoencoder reduces high-dimensional malware feature representations into a lower-dimensional latent space, addressing the inefficiency of distance metrics in high-dimensional settings, commonly referred to as the \textit{curse of dimensionality}.

The triplet loss is formulated to encourage semantic structure in the latent space. Each training instance is structured as a triplet: an anchor sample \( x_a \), a positive sample \( x_p \) from the same class, and a negative sample \( x_n \) from a different class. The objective is to ensure that the distance between the anchor and the positive is smaller than that between the anchor and the negative by at least a margin \( \alpha \), as defined by:

\begin{equation}
    L(x_a, x_p, x_n) = \max \left\{ 0, \, D(x_a, x_p) - D(x_a, x_n) + \alpha \right\}
\end{equation}

where

\begin{equation}
D(x_i, x_j) = \| f(x_i) - f(x_j) \|^2
\end{equation}

is the squared Euclidean distance between the embeddings produced by the encoder function \( f \). Over a batch of \( N \) triplets, the total triplet loss is given by:

\begin{equation}
L_{\text{triplet}} = \frac{1}{N} \sum_{i=1}^{N} L(x_{a_i}, x_{p_i}, x_{n_i})    
\end{equation}

In parallel, the MSE loss ensures the fidelity of the reconstruction produced by the decoder \( f' \). This encourages the network not only to learn discriminative embeddings but also to retain the structural integrity of the input features. The MSE loss between the original input \( x \) and its reconstruction \( \hat{x} = f'(f(x)) \) is defined as:

\begin{equation}
  L_{\text{MSE}} = \frac{1}{N} \sum_{i=1}^{N} \| x_i - \hat{x}_i \|^2  
\end{equation}

The combined loss function $L_{Final}$, incorporating both triplet and reconstruction objectives, guides the network to produce embeddings that are discriminative for classification while remaining informative enough for reconstruction \cite{guldemir2024addressing}.

\begin{equation}
    L_{Final}=L_{triplet}+\lambda L_{\text{MSE}}
\end{equation}

\subsection{Drift detection phase} \label{sec:drift-detection-setup}
Once the samples have been projected into the latent space via the triplet autoencoder, we perform unsupervised clustering to group samples that exhibit similar semantic and structural characteristics. This step is particularly important, as it allows us to not only cluster known malware families but also uncover meaningful substructures or behavioral variations within each family.

\begin{figure}[ht]
    \centering
    \includegraphics[width=0.8\linewidth]{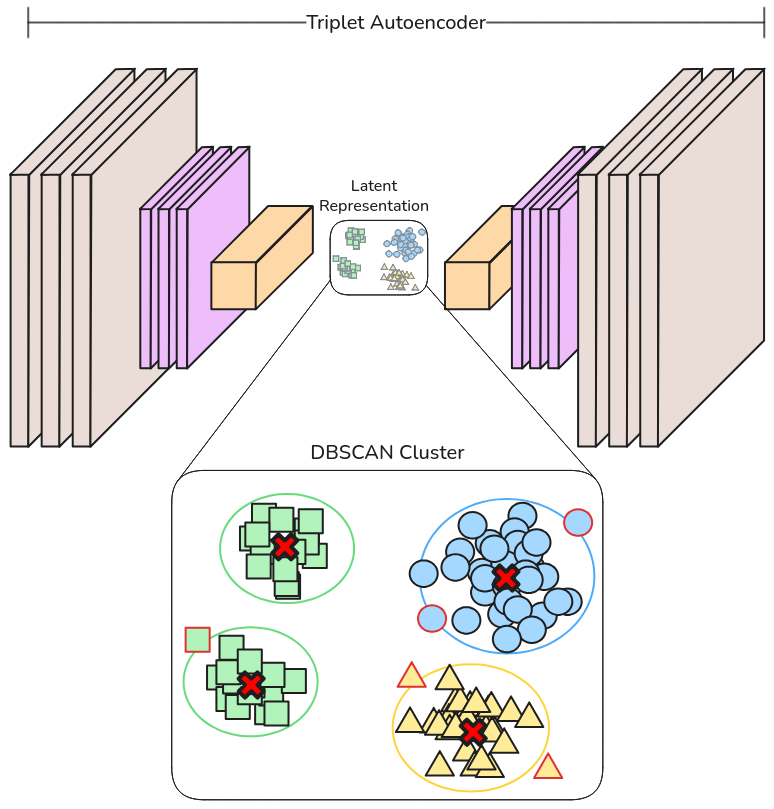}
    \caption{DBSCAN-derived subclusters and centroid-based acceptance thresholds in FARM.}
    \label{fig:triplet-dbscan}
\end{figure}

Rather than relying on traditional clustering techniques such as K-means, which require the number of clusters to be specified a priori, we employ the Density-Based Spatial Clustering of Applications with Noise (DBSCAN) algorithm \cite{ester1996density}. The primary motivation for using DBSCAN in our framework is its ability to identify an unknown number of dense subclusters directly from the embedding space, without requiring a fixed subgroup count for each malware family. This is particularly important in our setting, where a single malware family may exhibit multiple dense subclusters in the latent space due to internal behavioral diversity. Furthermore, DBSCAN distinguishes between core samples, border points, and noise, which is useful for reducing the influence of sparse or atypical samples during cluster construction. Prior studies have also reported strong performance of DBSCAN in malware clustering contexts \cite{faridi2018performance, kinable2011malware}, further supporting its use in our framework. Accordingly, DBSCAN is adopted here as a practical mechanism for discovering family-level substructure under unknown complexity in the latent space. This clustering process is illustrated in Figure~\ref{fig:triplet-dbscan}, where samples are embedded using a triplet autoencoder and then grouped via DBSCAN.

DBSCAN relies on two key hyperparameters: the neighborhood radius \( \varepsilon \), which defines how close points need to be to be considered part of a cluster, and the minimum number of points \( minPts \) required to form a dense region. A commonly used heuristic for setting \( minPts \) is to take twice the number of dimensions in the latent space, i.e., \( minPts = 2 \times \text{dim} \) \cite{schubert2017dbscan}. To determine a suitable value for \( \varepsilon \), we utilize the k-distance plot method. This involves plotting the distance to each point’s k-th nearest neighbor (where \( k = minPts \)) in ascending order. A distinct \textit{elbow} or \textit{knee} in the curve typically indicates a good value for \( \varepsilon \), as it marks the transition between dense regions (clusters) and sparse regions (noise).

Once clustering is performed using DBSCAN, we proceed to calculate a threshold value for each cluster to enable classification and outlier detection during inference. Importantly, these clusters do not necessarily correspond one-to-one with malware families; a single family may yield multiple clusters due to internal variation or subfamily structure present in the training data. The threshold associated with each cluster defines the maximum acceptable distance between a sample and its assigned cluster's centroid, and is used to determine whether a new sample belongs to a known family or should instead be flagged as drifted. 

Let \( C = \{ C_1, C_2, \ldots, C_k \} \) be the set of clusters returned by DBSCAN, where each cluster \( C_i \) contains \( n_i \) samples. Let \( f(x) \in \mathbb{R}^d \) denote the embedding of a sample \( x \) in the latent space, as computed by the encoder \( f \).

For each cluster \( C_i \), we first compute the cluster centroid \( \mu_i \) as the mean of the embedded vectors in that cluster:

\begin{equation}    
\mu_i = \frac{1}{n_i} \sum_{x \in C_i} f(x)
\end{equation}

Next, for each sample \( x \in C_i \), we compute the squared Euclidean distance between the sample’s embedding and its cluster centroid:

\begin{equation}
d_x = \| f(x) - \mu_i \|^2
\end{equation}

The threshold \( \tau_i \) for cluster \( C_i \) is defined as the maximum distance between any in-cluster sample and the cluster centroid:

\begin{equation}
\tau_i = \max_{x \in C_i} \| f(x) - \mu_i \|^2
\end{equation}

This threshold \( \tau_i \) defines a cluster-specific acceptance boundary around the centroid of \( C_i \), determined by the maximum distance of its non-outlier members in the embedding space. During inference, a new sample \( x_{\text{new}} \) is mapped to \( f(x_{\text{new}}) \) and assigned to the cluster with the nearest centroid only if:

\begin{equation}
\| f(x_{\text{new}}) - \mu_i \|^2 \leq \tau_i
\end{equation}

Otherwise, it is rejected as an outlier and flagged for further analysis. This strategy provides a lightweight, data-driven approximation for cluster assignment that is consistent with the metric space learned by the triplet autoencoder.

\subsection{Drift adaptation phase}

The drift adaptation phase addresses the emergence of novel malware families not represented in the initial training distribution. Instead of retraining the model upon the arrival of each new class, the system adopts a two-stage strategy that balances rapid response with long-term integration. This design targets drift-aware adaptation under evolving malware behavior, where rapid response and later consolidation are both important, but it is not intended to constitute a full classical continual learning framework with long sequential task updates and explicit forgetting-aware optimization. Initially, EMBER features extracted from a PE malware are embedded into a latent space using the triplet encoder. If the sample is identified as drifted, meaning it exceeds the threshold of all known clusters, it is stored in a buffer for further processing. Upon each addition to the buffer, DBSCAN clustering is re-applied to the accumulated drifted samples. Unlike the clustering used during the initial drift detection phase, DBSCAN is applied here to discover cohesive groupings within the buffered samples. The minimum cluster size is set to 10, which serves as a practical compromise between robustness and early adaptation: smaller groups are more likely to be unstable, while larger requirements would delay recognition of emerging malware families. This choice is consistent with the few-shot analysis in Section~\ref{sec:n-way-k-shot}, where performance improves with additional support samples, making 10 a reasonable balance between accuracy and responsiveness. Once such a cluster is found, its mean embedding is computed to form a new prototype, which is then added to the set of known clusters. This enables future samples to be classified into the newly identified group, thereby allowing the model to adapt to novel malware families without full retraining. To distinguish between representations of known and novel families, we refer to the known cluster means as centroids and to the novel representations as prototypes. Although both are computed as class-wise means, this terminological distinction reflects their respective roles within the system.

While our method does not rely on a dedicated prototypical network architecture \cite{snell2017prototypical}, the adaptation mechanism is functionally similar, enabling the model to recognize new families with only minimal labeled data. In the longer term, once sufficient samples have been collected, the model is retrained to fully incorporate the new class into the embedding space and clustering structure.

\paragraph{Few-shot adaptation with prototype-based inference.}
Our approach leverages the triplet-trained encoder \( f : \mathbb{R}^d \rightarrow \mathbb{R}^m \) to project samples into a latent space where semantically similar inputs are placed closer together. Afterwards, we perform classification by comparing query samples to class prototypes—mean embeddings computed from labeled support examples.

Given a small support set \( \mathcal{S} = \{(x_i, y_i)\}_{i=1}^{N \cdot K} \), where \( N \) is the number of classes and \( K \) is the number of examples per class, we compute the prototype for class \( n \) as:

\begin{equation}
    \mathbf{p}_n = \frac{1}{K} \sum_{i: y_i = n} f(x_i)
\end{equation}

A query sample \( x_q \) is classified by assigning it the label of the nearest prototype in Euclidean space:

\begin{equation}
\hat{y}_q = \arg\min_{n} \| f(x_q) - \mathbf{p}_n \|^2
\end{equation}

This inference procedure allows rapid adaptation to novel classes using only a few labeled examples, without retraining the encoder.

\paragraph{Transition to full retraining.}
While few-shot adaptation provides an immediate mechanism for recognizing novel malware families, its effectiveness remains constrained by the limited number of initial support samples. To enable long-term integration, the system transitions to full retraining once a sufficient number of confidently classified samples has been accumulated. These samples are not manually labeled but are instead collected automatically based on their assignment to the novel family via the prototype introduced during the adaptation phase.

As described previously, when a cohesive cluster of drifted samples is detected, a prototype is formed and incorporated into the set of known malware family representations. This prototype is then used during inference to classify incoming samples. The system continuously monitors these assignments and accumulates samples that are classified into the novel family. Once the number of such samples reaches a threshold sufficient to support stable retraining, the model undergoes a full retraining procedure.

During this phase, the triplet autoencoder is updated using both the original training data and the newly accumulated samples from the novel malware family. Following retraining, all embeddings are recomputed, and updated centroids and thresholds are established. The previously used prototype is thereby replaced with a centroid derived from the retrained embedding space, formally integrating the new family into the model’s representation and classification framework.

This retraining strategy, based on inference-driven sample accumulation, ensures both rapid adaptation to novel threats and consistent performance over time in the presence of evolving malware distributions.

\section{Evaluation and Experimental Setup} \label{sec:eval}
To assess the effectiveness of FARM under dynamic malware scenarios, we structure our evaluation around the following research questions:

\begin{itemize}
    \item \textbf{RQ1:} \textit{How effectively can metric learning, particularly through a triplet autoencoder, detect concept drift in malware classification?} This question is addressed in \autoref{sec:drift-detection}.
    
    \item \textbf{RQ2:} \textit{Can few-shot learning using prototype-based classification accurately adapt to both evolved and entirely new malware families with only a small number of labeled samples?} This question is addressed in \autoref{sec:drift-adaptation}.
    
    \item \textbf{RQ3:} \textit{To what extent can FARM adapt to previously unseen malware families under label drift using only a small number of labeled samples?} This question is addressed in \autoref{sec:adaptation-label-drift}
\end{itemize}

\subsection{Dataset Description}

For our experiments, we use the BenchMFC dataset \cite{jiang2024benchmfc} as the source benchmark. BenchMFC comprises over 223{,}000 unpacked malware samples spanning 526 known families, collected between 2012 and 2022. Each sample is annotated with its malware family and a timestamp indicating its earliest known appearance, enabling temporal segmentation of malware evolution. The dataset covers a broad range of malware types and families, making it suitable for evaluating classification systems under distributional change.

In our setup, we focus exclusively on unpacked samples and do not consider packing-related drift or packed variants, as packing can significantly impact concept drift and has been shown to challenge the generalization ability of static machine learning-based detectors \cite{aghakhani2020malware}. We leave the exploration of such effects to future work. Following \cite{jiang2024benchmfc}, we adopt the same controlled 16-family subset summarized in \autoref{tab:dataset_composition}. This subset is organized into three categories:

\begin{itemize}
    \item known families used for training,
    \item evolving families used for covariate drift evaluation, and
    \item unseen families used for label drift evaluation.
\end{itemize}

Both the malware family labels and the drift categorization are adopted directly from the original BenchMFC dataset, with no additional labeling applied. Each category contains eight malware families, with 500 Windows PE malware samples per family, forming a balanced dataset for evaluating metric learning and adaptation strategies under both covariate and label drift scenarios.

\begin{table*}[ht]
    \centering
    \caption{Dataset composition.}
    \label{tab:dataset_composition}
    \renewcommand{\arraystretch}{1.25}
    \setlength{\tabcolsep}{8pt}
    \begin{tabular}{@{} l r l r l r @{}}
        \toprule
        \multicolumn{2}{c}{\textbf{Training/Validation}} &
        \multicolumn{2}{c}{\textbf{Evolved Evaluation}} &
        \multicolumn{2}{c}{\textbf{Unseen Evaluation}} \\
        \cmidrule(lr){1-2}\cmidrule(lr){3-4}\cmidrule(lr){5-6}
        \textbf{Family} & \textbf{Size} &
        \textbf{Family} & \textbf{Size} &
        \textbf{Family} & \textbf{Size} \\
        \midrule
        simda    & 500 & simda    & 500 & hupigon     & 500 \\
        gandcrab & 500 & gandcrab & 500 & wannacry    & 500 \\
        yuner    & 500 & yuner    & 500 & vobfus      & 500 \\
        hotbar   & 500 & hotbar   & 500 & imali       & 500 \\
        fareit   & 500 & fareit   & 500 & onlinegames & 500 \\
        zbot     & 500 & zbot     & 500 & lydra       & 500 \\
        upatre   & 500 & upatre   & 500 & zlob        & 500 \\
        parite   & 500 & parite   & 500 & virut       & 500 \\
        \midrule
        \textbf{Total} & \textbf{4000} &
        \textbf{Total} & \textbf{4000} &
        \textbf{Total} & \textbf{4000} \\
        \bottomrule
    \end{tabular}
\end{table*}

The BenchMFC dataset provides both the raw PE malware binaries and their corresponding EMBER~\cite{anderson2018ember} feature representations. As EMBER is a widely used and well-established feature set in static malware classification~\cite{yang2021bodmas, harang2020sorel, jiang2024benchmfc}, we adopt it in our experiments to represent PE malware samples. EMBER encodes a wide range of structural and semantic characteristics of executable files, including byte histograms, PE header information, section metadata, and imported libraries. These high-dimensional features provide a comprehensive and consistent representation across samples, making them well-suited for training discriminative embedding models and evaluating performance under distributional shift.

\subsection{Triplet Autoencoder Training} \label{sec:training-setup}

To learn a discriminative latent space for malware representation, we train a triplet autoencoder model using the malicious samples in our selected subset of the BenchMFC dataset. These samples correspond to early, unpacked instances from known families and do not include evolved or packed variants. We reserve 80\% of these 4000 samples for training and use the remaining 20\% for validation.

Before training, we apply a Quantile Transformer to normalize the input features, as the raw data exhibits extreme variation in scale, ranging from small decimals to values in the millions. We follow this with variance thresholding to remove near-constant features. These preprocessing steps reduce the input dimensionality from 2381 to 1830, resulting in a more stable and informative feature set.

The model is built as a symmetrical encoder-decoder architecture. The encoder progressively reduces the 1830-dimensional input to a 32-dimensional embedding using three fully connected layers of sizes 1024, 256, and 32, each followed by batch normalization, \textit{ReLU} activation and dropout for regularization. The decoder mirrors this structure in reverse, expanding the 32-dimensional embedding back to the original input space through layers of sizes 256 and 1024, culminating in a reconstruction layer of size 1830.

Although a typical triplet network only requires the encoder, we employ a full autoencoder setup to incorporate reconstruction loss alongside the triplet loss. This dual-objective training encourages the model to learn embeddings that not only preserve semantic similarity among malware samples but also retain sufficient information for reconstructing the input. The combined loss function prioritizes discriminative power by weighting the triplet loss twice as much as the reconstruction loss, using a $\lambda=0.5$ during training.

\subsection{DBSCAN Clustering and Thresholding Setup}
\label{dbscan-setup}

For each known malware family in the training set, we independently apply DBSCAN clustering in the learned embedding space. The neighborhood radius parameter \( \varepsilon \) is selected per class by analyzing the sorted distances to each point’s \( k \)-th nearest neighbor. The optimal value is chosen at the point where the curve exhibits a noticeable inflection, commonly referred to as the “elbow”, which marks the transition from dense to sparse regions. The minimum number of samples parameter is fixed based on the dimensionality of the embedding space.

After clustering, only the inlier samples, those assigned to valid DBSCAN clusters, are used to compute thresholds. For each cluster, we calculate the centroid and measure the Euclidean distance from each inlier to its centroid. A threshold is then defined as the mean of these distances plus a multiple of their standard deviation. These thresholds are later used during inference to determine whether a test sample should be accepted as belonging to a known class or rejected as a potential outlier.

\section{Evaluation: Drift Detection} \label{sec:drift-detection}

In this section, we evaluate the FARM framework's ability to detect concept drift in Windows PE malware classification. We focus on two distinct forms of drift: \textit{covariate drift}, which involves changes in the behavior of known malware families, and \textit{label drift}, which corresponds to the emergence of entirely new, previously unseen families. The goal of this stage is to determine whether FARM can accurately identify drifted samples before any model adaptation takes place.

To provide a consistent reference point across all evaluation scenarios, we report the baseline performance of the trained FARM model on a held-out testing set comprising samples from the original training malware families. This set does not include any drifted or unseen variants and is evaluated using the model before any adaptation or retraining steps. Since this test set remains fixed and unaffected by concept drift, its results serve as a performance anchor for comparing the impact of covariate and label drift in subsequent experiments. Table \ref{tab:baseline-testing-set-performance} summarizes the per-family classification performance in terms of precision, recall, and F1 score.

\begin{table}[ht]
    \centering
    \caption{Baseline classification performance of the FARM on a testing set of known malware families.}
    \label{tab:baseline-testing-set-performance}
    \renewcommand{\arraystretch}{1.25}
    \setlength{\tabcolsep}{8pt}
    \begin{tabular}{@{} c c c c @{}}
        \toprule
        \textbf{Family} &
        \textbf{Precision} &
        \textbf{Recall} &
        \textbf{F1 Score} \\
        \midrule
        fareit & 0.92 & 0.90 & 0.91 \\
        gandcrab & 1.00 & 0.98 & 0.99 \\
        hotbar & 1.00 & 1.00 & 1.00 \\
        parite & 1.00 & 0.97 & 0.98 \\
        simda & 0.99 & 0.99 & 0.99 \\
        upatre & 0.94 & 0.97 & 0.95 \\
        yuner & 1.00 & 1.00 & 1.00 \\
        zbot & 0.89 & 0.94 & 0.91 \\
        \midrule
        \textbf{Average}     & \textbf{0.97} & \textbf{0.97} & \textbf{0.97} \\
        \bottomrule
    \end{tabular}
\end{table}

\subsection{Covariate Drift Evaluation (Evolved Families)}

To assess the model’s performance under covariate drift, we evaluate its performance on the evolved samples of each known malware family. These samples represent newer variants of previously seen families, which may exhibit behavioral or structural deviations from their earlier counterparts. Although all samples in this evaluation set are labeled as drifted in the dataset, they correspond to existing malware families and therefore represent covariate drift rather than entirely novel classes.

In contrast to label drift, where all unseen families are expected to be detected as drift by design, covariate drift presents a more nuanced challenge. Since the malware family is already known, a sample may or may not be detected as drift depending on how significantly its characteristics deviate from what the model has learned. As a result, evaluating drift detection alone may not provide a complete picture.

Table~\ref{tab:evolved_drift_detection_grouped} summarizes the results by reporting whether each sample was detected as drift and whether it was classified into the correct family. Correct classifications without drift detection indicate successful generalization to evolved variants. Correct classifications that are detected as drift suggest conservative behavior, where the model still assigns the correct family despite recognizing distributional change. Incorrect classifications that are not detected as drift imply failure to detect meaningful drift, while incorrect classifications that are flagged as drift reflect both generalization failure and successful drift detection.

To more precisely assess classification behavior, we report several complementary metrics. The error rate measures the proportion of non-drifted samples that were misclassified, providing a focused view of classification performance when the model chooses to make a prediction. In contrast, the drift rate indicates the percentage of total samples that were detected as drift, independent of classification correctness. The accuracy metric reflects overall classification performance across the entire set of evolved samples, including those flagged as drifted. Although classification is only applied to non-drifted samples, accuracy is reported to provide a general sense of how well the model performs under covariate drift. For example, in the case of the \texttt{fareit} family, a total of 500 evolved samples were evaluated. Among these, 333 were not detected as drift and were subsequently classified—all of which were correctly assigned to their original family. This yields an error rate of 0\%, indicating perfect classification among the samples the model attempted to classify. The remaining 167 samples were detected as drift, contributing to a drift rate of 33.4\%. Since accuracy is computed over the entire set of 500 samples, including those not classified due to being flagged as drift, the overall accuracy for \texttt{fareit} appears is 74\%.

Overall, the model achieves 81.4\% accuracy on evolved families, with 31.9\% of the samples detected as drift. Among samples not detected as drift, the error rate remains relatively low at 5.8\%. The model generalizes particularly well to families such as \texttt{hotbar}, \texttt{parite}, and \texttt{yuner}, while showing higher rates of drift detection and misclassification in families like \texttt{simda} and \texttt{upatre}, indicating more pronounced behavioral evolution in those cases.

\begin{table*}[ht]
    \begin{center}  
    \caption{Drift detection results on evolved families, grouped by drift decision and classification correctness.}
    \label{tab:evolved_drift_detection_grouped}
    \renewcommand{\arraystretch}{1.25}
    \scriptsize
    \setlength{\tabcolsep}{8pt}
    \begin{tabular}{@{} l c c c c c c c c @{}}
        \toprule
        & & \multicolumn{2}{c}{\textbf{Not Drifted}} & \multicolumn{2}{c}{\textbf{Drifted}} & & & \\
        \cmidrule(lr){3-4}\cmidrule(lr){5-6}
        \textbf{Family} & \textbf{\# of Samples} &
        \textbf{Correct} & \textbf{Wrong} &
        \textbf{Correct} & \textbf{Wrong} &
        \textbf{\shortstack{Rejection \\ Rate}} &
        \textbf{\shortstack{Accuracy \\ Error Rate}} &
        \textbf{Accuracy} \\
        \midrule
        fareit & 500 & 333 & 0 & 37 & 130 & 0.33 & 0.00 & 0.74 \\
        gandcrab & 500 & 314 & 12 & 123 & 51 & 0.35 & 0.02 & 0.87 \\
        hotbar & 500 & 489 & 0 & 11 & 0 & 0.02 & 0.00 & 1.00 \\
        parite & 500 & 463 & 0 & 34 & 3 & 0.07 & 0.00 & 0.99 \\
        simda & 500 & 141 & 104 & 134 & 121 & 0.51 & 0.21 & 0.55 \\
        upatre & 500 & 126 & 16 & 120 & 238 & 0.72 & 0.03 & 0.49 \\
        yuner & 500 & 500 & 0 & 0 & 0 & 0.00 & 0.00 & 1.00 \\
        zbot & 500 & 200 & 25 & 231 & 44 & 0.55 & 0.05 & 0.86 \\
        \midrule
        \textbf{Total} & \textbf{4000} & \textbf{2566} & \textbf{157} & \textbf{690} & \textbf{587} & \textbf{0.32} & \textbf{0.06} & \textbf{0.81} \\
        \bottomrule
    \end{tabular}
    \end{center}
\end{table*}

\subsection{Label Drift Evaluation (Unseen Families)} \label{sec:label_shift_eval}

To evaluate the system's ability to detect unseen samples, we assess its performance on samples from malware families that were entirely excluded during training. These unseen families simulate a label drift scenario, where new classes emerge that are not part of the known threat taxonomy. The model is expected to detect such drifts by rejecting these samples as unfamiliar or out-of-distribution.

Table~\ref{tab:label_shift_detection} presents per-family results, summarizing the rate at which the system identified and flagged these novel inputs. Overall, the model correctly identified 84\% of these samples as drifted, demonstrating strong performance in recognizing previously unseen families. However, detection performance varied across families. Notably, the \texttt{vobfus} family exhibited a substantially lower drift detection rate of only 20\%, indicating that the model frequently misclassified its samples as belonging to known families. This behavior can be attributed to the semantic and behavioral characteristics of \texttt{vobfus}. According to Microsoft Defender reports \cite{microsoft2025security}, \texttt{vobfus} commonly acts as a dropper for other malware families such as \texttt{fareit} and \texttt{zbot}, both of which are present in the training set. This functional overlap likely causes the encoder to position \texttt{vobfus} samples close to these known families in the embedding space, making them appear familiar and reducing the likelihood of being flagged as drifted.

\begin{table}[ht]
    \centering
    \caption{Drift detection results for unseen malware families (label drift scenario).}
    \label{tab:label_shift_detection}
    \renewcommand{\arraystretch}{1.25}
    \setlength{\tabcolsep}{8pt}
    \begin{tabular}{@{} c c c c c @{}}
        \toprule
        \textbf{Family} & 
        \textbf{\# of Samples} & 
        \textbf{Inliers} & 
        \textbf{Drifted} & 
        \textbf{Drift Rate} \\
        \midrule
        hupigon & 500 & 34 & 466 & 0.93 \\
        imali & 500 & 0 & 500 & 1.00 \\
        lydra & 500 & 24 & 476 & 0.95 \\
        onlinegames & 500 & 18 & 482 & 0.96 \\
        virut & 500 & 72 & 428 & 0.86 \\
        vobfus & 500 & 402 & 98 & 0.20 \\
        wannacry & 500 & 1 & 499 & 1.00 \\
        zlob & 500 & 83 & 417 & 0.83 \\
        \midrule
        \textbf{Total} & \textbf{4000} & \textbf{634} & \textbf{3366} & \textbf{0.84} \\
        \bottomrule
    \end{tabular}
\end{table}

\begin{figure*}[ht]
    \centering
    \includegraphics[width=\linewidth]{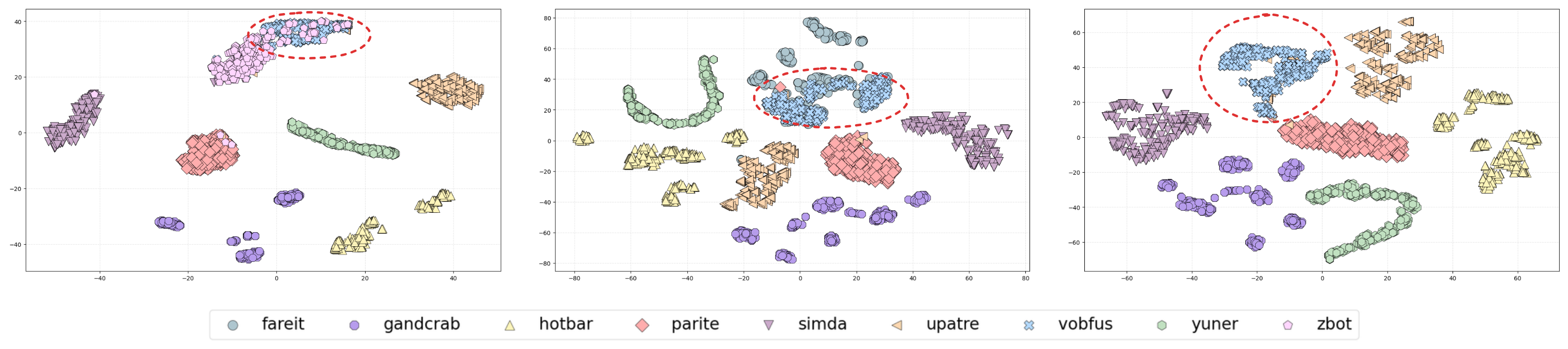}
    \caption{t-SNE visualizations of latent embeddings for \texttt{vobfus} and related families under three ablation configurations. Red ellipses highlight \texttt{vobfus} clusters. Left: model trained with \texttt{zbot} but excluding \texttt{fareit}. Middle: model trained with \texttt{fareit} but excluding \texttt{zbot}. Right: both \texttt{fareit} and \texttt{zbot} excluded.}
    \label{fig:vobfus_tsne}
\end{figure*}

To validate this hypothesis, we conducted ablation experiments by retraining the model under three configurations: (a) training with \texttt{zbot} included but excluding \texttt{fareit}, (b) training with \texttt{fareit} included but excluding \texttt{zbot}, and (c) excluding both \texttt{fareit} and \texttt{zbot}. Figure~\ref{fig:vobfus_tsne} shows t-SNE visualizations of the resulting embeddings, with \texttt{vobfus} clusters highlighted in red. In the first two configurations, \texttt{vobfus} samples align closely with the single remaining related family, suggesting that their behavioral similarity causes embedding overlap. Only after removing both families does \texttt{vobfus} form a distinct and separable cluster. These findings confirm that drift detection may fail when novel families share strong structural or behavioral features with previously seen malware.

Taken together, the results support RQ1, indicating that FARM effectively detects both types of concept drift through its metric learning and clustering approach.

\subsection{Comparison with Baselines}

To assess FARM under label drift, we compare it against CADE~\cite{yang2021cade} and a controlled ablated variant of our method. CADE and FARM both use an autoencoder-based architecture to learn a latent representation, but they differ in the metric-learning objective: CADE employs contrastive loss, which learns pairwise similarity constraints, whereas FARM uses triplet loss, which explicitly encourages relative separation between anchor, positive, and negative samples. In addition to this external comparison, we include the \textit{Percentile} variant to isolate the effect of the class-boundary modeling strategy. \textit{Percentile} uses the same triplet autoencoder and training setup as FARM, but defines class boundaries using class-wise mean centroids and fixed distance thresholds based on the 95th percentile. In contrast, FARM models each class using DBSCAN-derived clusters with cluster-specific dynamic thresholds. The triplet-based variants use the same autoencoder architecture and training settings described in \autoref{sec:training-setup}, enabling a controlled comparison of boundary-definition strategies.

\begin{table}[ht]
    \centering
    \caption{Comparison of classification performance on unseen families (label drift scenario). FARM shows a better balance between precision and recall.}
    \label{tab:method_comparison}
    \renewcommand{\arraystretch}{1.25}
    \setlength{\tabcolsep}{8pt}
    \begin{tabular}{@{} c c c c @{}}
        \toprule
        \textbf{Method} &
        \textbf{Precision} &
        \textbf{Recall} &
        \textbf{F1 Score} \\
        \midrule
        CADE       & 0.6395 $\pm$ 0.0181 & 0.9612 $\pm$ 0.0680 & 0.7676 $\pm$ 0.0356 \\
        Percentile & 0.7961 $\pm$ 0.1204 & 0.8130 $\pm$ 0.2569 & 0.7924 $\pm$ 0.2136 \\
        FARM       & 0.8372 $\pm$ 0.0936 & 0.8415 $\pm$ 0.2505 & 0.8254 $\pm$ 0.2031 \\
        \bottomrule
    \end{tabular}
\end{table}

Comparative results are presented in Table \ref{tab:method_comparison}. FARM achieves stronger overall performance than both CADE and the ablated \textit{Percentile} variant, providing a better balance between precision and recall. The comparison with CADE provides an empirical contrast between FARM’s triplet-loss-based latent representation and an autoencoder trained with contrastive loss, while the comparison with \textit{Percentile} isolates the effect of the boundary modeling strategy, since both methods share the same triplet-based representation. The results suggest that DBSCAN-based clustering with cluster-specific dynamic thresholds improves class boundary modeling and leads to better performance under label drift. The reported values correspond to the mean and standard deviation of precision, recall, and F1 score across experiments on all eight unseen malware families. In each run, one novel family is introduced at a time and evaluated independently.

We observe relatively high standard deviation values across methods, particularly in recall and F1 score. This variance is primarily attributed to the \texttt{vobfus} family, which exhibits inconsistent drift detection performance due to its behavioral similarity with previously seen families such as \texttt{fareit} and \texttt{zbot}. As discussed in \autoref{sec:label_shift_eval}, \texttt{vobfus} samples are frequently embedded near these known families in the latent space, making them harder to distinguish as novel. This leads to occasional misclassification and variability in detection performance across evaluation splits. When \texttt{vobfus} is excluded, the standard deviations for all metrics decrease substantially, confirming its strong influence on overall variance.

\section{Evaluation: Drift Adaptation} \label{sec:drift-adaptation}
\subsection{Experimental Setup}

In this section, we describe two key experiments designed to evaluate the FARM framework’s ability to adapt to both covariate and label drift. Both experiments utilize a unified adaptation pipeline, which consists of three main components: drift detection (as introduced in \autoref{sec:drift-detection}), incremental clustering of drifted samples, and the integration of class prototypes for classification.

Drift detection is carried out in the latent space produced by the triplet encoder. Each incoming malware sample is projected into this space and compared against the centroids of existing clusters. If the distance between the sample and its nearest cluster centroid exceeds a class-specific threshold, the sample is flagged as drifted. Drifted samples are not immediately used for classification; instead, they are accumulated in a buffer that stores candidate samples for potential cluster formation.

Clustering is initiated when the buffer reaches at least 10 drifted samples, aligning with the number of support examples used in our few-shot classification setting. At this point, the DBSCAN algorithm is applied to the buffered samples in the latent space. The \texttt{min\_samples} parameter is set to 10 to ensure that only sufficiently dense and coherent groupings are considered for prototype creation. For the \( \varepsilon \) parameter, we follow the same method described in \autoref{sec:drift-detection-setup}.  If DBSCAN fails to identify a valid cluster, new drifted samples continue to be added incrementally to the buffer, and clustering is reattempted with each new addition until a cluster of at least 10 samples is detected.

Once such a cluster is formed, its centroid is treated as a class prototype and incorporated into the classification model. In the case of label drift, where the cluster represents a previously unseen malware family, this prototype is treated as a novel class, allowing the system to classify new instances with only a few examples. In contrast, for covariate drift, where the cluster corresponds to behavioral or structural variations within a previously known malware family, the new prototype is linked to the corresponding existing family label in the training set. In other words, instead of creating a new class, the prototype is added as an additional representative of that known malware family, enriching its characterization in the embedding space. In both scenarios, classification is performed using nearest-prototype matching in the latent space.

\subsection{Adaptation under Covariate Drift}

To evaluate the system's ability to adapt to evolving behavior within known classes, we assess prototype expansion under covariate drift. For each evolving malware family, incoming samples are first evaluated against existing class prototypes using a threshold-based distance function. Samples exceeding their class-specific threshold are flagged as drifted and buffered for potential adaptation.

Once the buffer contains a sufficient number of drifted samples, we apply DBSCAN to identify cohesive latent clusters. If a cluster with at least $k=10$ samples is found, the mean of those embeddings is promoted as a prototype for the corresponding family. The classifier is then updated to use this extended prototype set and re-evaluated on both the testing set and evolved samples.

Table~\ref{tab:covariate_adaptation_results} presents the F1 scores with and without adaptation across the different evolving malware families. Adaptation leads to consistent improvements in classification performance, with particularly significant gains observed for \texttt{upatre} (+19.44\%), \texttt{fareit} (+11.6\%), and \texttt{simda} (+9.1\%). These improvements suggest that the method is particularly effective in scenarios with covariate drift. In contrast, families such as \texttt{hotbar}, \texttt{yuner}, and \texttt{parite} already had high F1 scores before adaptation, indicating minimal or no observable drift. Overall, the approach achieves an average improvement of 5.6\% in F1 score, demonstrating its broad effectiveness.

\begin{table}[ht]
    \centering
    \small
    \caption{Performance of drift adaptation on evolving malware families (covariate drift)}
    \label{tab:covariate_adaptation_results}
    \renewcommand{\arraystretch}{1.25}
    \setlength{\tabcolsep}{8pt}
    \begin{tabular}{@{} c c c c @{}}
        \toprule
        \textbf{Family} &
        \makecell{\textbf{F1 Score} \\ \textbf{(No Adaptation)}} &
        \makecell{\textbf{F1 Score} \\ \textbf{(Adaptation)}} &
        \textbf{Improvement (\%)} \\
        \midrule
        fareit     & 0.86 & 0.96 & +11.6 \\
        gandcrab   & 0.94 & 0.97 & +3.2  \\
        hotbar     & 1.00 & 1.00 & +0.0   \\
        parite     & 0.99 & 0.99 & +0.2  \\
        simda      & 0.77 & 0.84 & +9.1 \\
        upatre     & 0.72 & 0.86 & +19.44 \\
        yuner      & 1.00 & 1.00 & +0.0   \\
        zbot       & 0.92 & 0.94 & +2.17  \\
        \midrule
        \textbf{Average} & \textbf{0.90} & \textbf{0.95} & \textbf{+5.6} \\
        \bottomrule
    \end{tabular}
\end{table}

\subsection{Adaptation under Label Drift} \label{sec:adaptation-label-drift}
To evaluate the system's ability to adapt to label drift, we simulate the emergence of a novel malware family that were entirely excluded from the training phase. Samples from each unseen family are introduced one at a time. When a sample is identified as drifted, based on its distance from all known class prototypes, it is stored in a buffer designated for potential novel class discovery. Once the buffer accumulates enough drifted samples, DBSCAN is applied in the latent space to identify cohesive clusters. If a valid cluster is detected, its mean embedding is used to form a new class prototype, which is then incorporated into the classifier. This enables the system to recognize future samples from that family without requiring full model retraining.

\begin{table}[ht]
    \centering
    \caption{Few-shot adaptation results for individual unseen malware families under label drift. Each family was excluded during training and later reintroduced in the test stream for drift detection and adaptation.}
    \label{tab:drift_adaptation_results}
    \renewcommand{\arraystretch}{1.25}
    \setlength{\tabcolsep}{8pt}
    \begin{tabular}{@{} c c c c @{}}
        \toprule
        \textbf{Family} &
        \textbf{Precision} &
        \textbf{Recall} &
        \textbf{F1 Score} \\
        \midrule
        hupigon     & 0.99 & 0.63 & 0.77 \\
        imali       & 0.99 & 0.97 & 0.98 \\
        lydra       & 0.99 & 0.93 & 0.96 \\
        onlinegames & 0.99 & 0.76 & 0.86 \\
        virut       & 0.98 & 0.73 & 0.84 \\
        vobfus      & 0.94 & 0.45 & 0.60 \\
        wannacry    & 0.96 & 0.96 & 0.96 \\
        zlob        & 0.98 & 0.76 & 0.86 \\
        \midrule
        \textbf{Average}     & \textbf{0.98} & \textbf{0.77} & \textbf{0.85} \\
        \bottomrule
    \end{tabular}
\end{table}

Table~\ref{tab:drift_adaptation_results} presents precision, recall, and F1 score for each of the eight unseen malware families after few-shot adaptation. FARM achieves an average F1 score of 0.85 across the novel families, indicating effective integration of previously unseen classes using minimal supervision. High F1 scores for families like \texttt{imali}, \texttt{lydra}, and \texttt{wannacry} suggest well-formed, separable clusters in the latent space. In contrast, lower scores for \texttt{vobfus} and \texttt{hupigon} reflect weaker adaptation, likely due to ambiguous embeddings. As previously discussed in~\autoref{sec:label_shift_eval}, \texttt{vobfus} in particular shows substantial overlap with known families such as \texttt{fareit} and \texttt{zbot}, which complicates drift detection and prototype formation.

In summary, these results address RQ2, demonstrating that FARM can successfully adapt to previously unseen malware families using only a small number of labeled examples, leveraging few-shot learning without retraining the underlying model.

\subsubsection{Generalization accuracy on unseen classes}
\label{sec:n-way-k-shot}

To evaluate generalization to previously unseen classes, we conduct standard few-shot classification experiments under multiple \(N\)-way \(K\)-shot settings. This evaluation protocol is widely used in the few-shot learning literature to assess whether a model can discriminate among \(N\) novel classes given only \(K\) labeled examples per class \cite{snell2017prototypical, chai2022dynamic, wang2021novel}.

In our setting, each \(N\)-way episode is formed by randomly selecting \(N\) unseen malware families from the eight families excluded during training. For each family, \(K\) labeled samples are used as the support set, and 15 additional samples are used as queries. This protocol reflects a realistic deployment scenario in which the model receives only a limited number of labeled samples from emerging threats and must generalize quickly to classify related variants. Following \cite{snell2017prototypical}, we run 600 episodes for each \(N\)-way \(K\)-shot configuration. The mean classification accuracy and 95\% confidence intervals reported in Table~\ref{tab:fewshot-results} are computed over these episodes, providing a statistically grounded estimate of generalization performance across tasks of varying difficulty.

As shown in Table~\ref{tab:fewshot-results}, performance improves consistently as the number of support samples per class increases. The largest gains are observed when moving from 1-shot to 5-shot, while the improvements from 10-shot to 20-shot are more modest, indicating diminishing returns beyond 10 labeled samples per class. In contrast, increasing the number of classes makes the task more difficult, leading to lower accuracy as the model must form finer-grained decision boundaries among a broader set of unseen families.

These results address RQ3 by showing that the learned embedding generalizes well to unseen malware families under limited-label conditions, while also indicating that around 10 support samples already provide a practical tradeoff between adaptation quality and labeling cost.

\begin{table}[ht]
    \centering
    \caption{Few-shot classification accuracy (\% ± 95\% confidence interval) across different N-way K-shot configurations.}
    \label{tab:fewshot-results}
    \renewcommand{\arraystretch}{1.25}
    \setlength{\tabcolsep}{8pt}
    \begin{tabular}{@{} c c c c c @{}}
        \toprule
        \textbf{N-way} &
        \textbf{1-shot} &
        \textbf{5-shot} &
        \textbf{10-shot} &
        \textbf{20-shot} \\
        \midrule
        3-way & 78.77 ± 1.26 & 86.16 ± 0.74 & 87.54 ± 0.64 & 88.40 ± 0.61 \\
        5-way & 70.64 ± 0.91 & 79.62 ± 0.63 & 81.24 ± 0.55 & 82.44 ± 0.53 \\
        8-way & 64.90 ± 0.67 & 74.53 ± 0.38 & 76.83 ± 0.31 & 77.79 ± 0.30 \\
        \bottomrule
    \end{tabular}
\end{table}

\subsection*{D. Retraining for Label Drift Integration}

While few-shot adaptation enables rapid incorporation of novel malware families, we also include a retraining strategy to support longer-term model integration. This mechanism is applied only for label drift, since the number of samples flagged under covariate drift is typically too small to support reliable model updates. To reduce the risk of overfitting to a very small set of newly observed samples, retraining is triggered only after the drift buffer accumulates at least 100 samples from a novel class. This threshold is used as a practical minimum sample budget for updating the encoder, so that retraining is not initiated from only a handful of potentially unrepresentative instances. In this sense, the threshold reflects a trade-off between adaptation speed and update stability: smaller buffers allow earlier response but may produce unstable latent-space adjustments, whereas larger buffers provide more reliable class representation at the cost of delayed integration. Although the value of 100 is empirically selected, it is intended as a conservative operating point rather than a fixed requirement, and can be adjusted depending on dataset scale, model capacity, and observed drift frequency. Once this threshold is reached, the model is retrained on the union of the original training data and the buffered drift samples, after which DBSCAN is reapplied to rebuild cluster structure and update the corresponding decision thresholds.


Table~\ref{tab:retrain_comparison} compares the performance of the few-shot adaptation and the retrained model across multiple unseen malware families. The retrained model consistently improves generalization, boosting the average F1 score from 0.85 to 0.94, a relative improvement of 10.6\%. This highlights the complementary nature of few-shot adaptation for rapid response and full retraining for deeper integration into the classification system.

\begin{table*}[ht]
    \centering
    \caption{Comparison of few-shot adaptation and full retraining on unseen malware families (label drift scenario).}
    \label{tab:retrain_comparison}
    \renewcommand{\arraystretch}{1.25}
    \setlength{\tabcolsep}{8pt}
    \begin{tabular}{@{} c c c c c c c @{}}
        \toprule
         &
        \multicolumn{3}{c}{\textbf{Few-shot Adaptation}} &
        \multicolumn{3}{c}{\textbf{Retraining}} \\
        \cmidrule(lr){2-4}\cmidrule(lr){5-7}
        \textbf{Family} & \textbf{Precision} & \textbf{Recall} & \textbf{F$_1$ Score}
        & \textbf{Precision} & \textbf{Recall} & \textbf{F$_1$ Score} \\
        \midrule
        hupigon        & 0.98 & 0.63 & 0.77 & 1.00 & 0.72 & 0.84 \\
        imali          & 0.99 & 0.97 & 0.98 & 1.00 & 0.99 & 1.00 \\
        lydra          & 0.97 & 0.94 & 0.95 & 1.00 & 0.93 & 0.97 \\
        onlinegames    & 0.97 & 0.77 & 0.86 & 1.00 & 0.96 & 0.98 \\
        virut          & 0.97 & 0.73 & 0.83 & 1.00 & 0.69 & 0.82 \\
        vobfus         & 0.94 & 0.45 & 0.61 & 1.00 & 0.98 & 0.99 \\
        wannacry       & 0.97 & 0.97 & 0.96 & 1.00 & 0.97 & 0.98 \\
        zlob           & 0.97 & 0.77 & 0.86 & 1.00 & 0.90 & 0.94 \\
        \midrule
        \textbf{Average} & \textbf{0.97} & \textbf{0.78} & \textbf{0.85} & \textbf{1.00} & \textbf{0.89} & \textbf{0.94} \\
        \bottomrule
    \end{tabular}
\end{table*}

\section{Discussion} \label{sec:discussion}

FARM shows that combining triplet-based metric learning with unsupervised clustering and few-shot adaptation can provide an effective mechanism for handling both covariate and label drift in malware classification. At the same time, several limitations of the current design should be noted. A central component of the system is DBSCAN clustering in the latent space, which enables the discovery of coherent malware subgroups without requiring predefined family counts. However, clustering performance remains sensitive to the selection of hyperparameters such as \( \varepsilon \) and minPts. In this work, \( \varepsilon \) was determined using a k-distance elbow heuristic \cite{schubert2017dbscan}, and minPts was set proportional to the latent dimension. While this provides a practical and automated calibration procedure, more adaptive or data-driven parameter selection could further improve drift detection sensitivity, particularly in noisy or structurally complex regions of the latent space.

An illustrative example of this sensitivity is the \texttt{vobfus} family, which exhibited lower drift detection accuracy due to behavioral overlap with training-set families such as \texttt{fareit} and \texttt{zbot}. This overlap caused \texttt{vobfus} samples to be embedded near existing families, resulting in frequent misclassifications. Such cases highlight a limitation of the current embedding space in separating semantically similar but distinct malware behaviors. Representation quality may also be affected by the construction of training triplets. In the current approach, anchor-positive-negative combinations are sampled randomly, without using strategies such as hard positive or hard negative mining. Although this simplifies training and avoids overemphasizing extreme cases, it may limit the model’s ability to refine decision boundaries in ambiguous or tightly clustered regions.

A further limitation concerns the evaluation protocol. FARM is assessed here in a drift-response setting, where the model is first trained on an initial family distribution and then updated through prototype insertion and later retraining as drifted samples accumulate. While this setup reflects an operational malware adaptation scenario and includes a longer-term consolidation stage, it does not constitute a full continual learning benchmark with many sequential tasks, replay strategies, or explicit forgetting measurements. Therefore, the present results should be interpreted as evidence of effective drift-aware adaptation rather than as a claim of superiority over established continual learning methods.

Another scope limitation is that the present study is instantiated on EMBER feature representations for Windows PE malware. Although the overall framework is compatible with other feature extractors, the quality of the learned latent space and the resulting drift boundaries will depend on the representational properties of the chosen features. Extending the framework to alternative static, dynamic, or hybrid malware representations remains an important direction for future work.

Overall, the results suggest that FARM provides a practical basis for drift-aware malware adaptation, while also revealing several areas where the current formulation can be strengthened. In particular, improving clustering calibration, refining triplet construction, and evaluating the framework under broader sequential adaptation settings would further clarify its robustness and generality.

\section{Conclusion} \label{sec:conclusion}
This work addresses the challenge of maintaining malware family classification performance under continuous concept drift. By integrating metric learning, few-shot adaptation, and conditional retraining, FARM provides a unified framework for identifying and responding to both evolved and novel malware families. The use of a triplet autoencoder enables drift detection without relying on softmax confidence scores, while few-shot learning supports rapid adaptation using only a small number of labeled samples. In addition, the inclusion of a buffer-triggered retraining mechanism allows FARM to incorporate newly observed data into the model and refresh its latent representation for longer-term updating. Experimental results show that FARM improves classification performance across both covariate drift and label drift scenarios, indicating its effectiveness in temporally evolving settings. These findings suggest that FARM can serve as a practical approach for drift-aware malware family classification in dynamic environments. Future work may extend the framework to handle packed malware, which introduces additional drift through obfuscation and poses challenges for static analysis. Another important direction is to improve the interpretability of drift decisions by identifying the underlying changes that trigger detection, thereby increasing transparency and supporting more informed operational responses.

\vspace{0.5em}

\noindent\textbf{Code Availability:} The source code and scripts developed for this paper are available at \url{https://github.com/numanhg/farm}.

\printbibliography

\end{document}